\documentclass[preprint,12pt]{elsarticle}

\usepackage{amssymb}
\usepackage{amsmath}

\usepackage{placeins} % For the float barrier
\usepackage[acronym]{glossaries} % For abbreviations because fuck this, use \gls{} to ensure full print on first use, use acrlong if long version is desired
\usepackage{tcolorbox} % For comments
\usepackage{soul} % For the highlight in the notes
\usepackage{siunitx} % for units and spacing of numbers
\usepackage{chemformula} % Formula subscripts using \ch{}
\usepackage{chemmacros} % For the one name
\usepackage[colorlinks=true]{hyperref} % remove before submitting?
\usepackage{cleveref} % please god allow me to use this

\newacronym{ftir}{FTIR}{fourier-transform infrared spectroscopy}
\newacronym{tr-uv-vis}{TR UV-VIS}{time resolved UV-VIS spectroscopy}
\newacronym{tof-sims}{ToF-SIMS}{time-of-flight secondary ion mass spectrometry}
\newacronym{pv}{PV}{photovoltaic}
\newacronym[plural=pv-photo,firstplural=photovoltaic (PV) photographs]{pv-photo}{PV photograph}{photovoltaic photograph}
\newacronym{dssc}{DSSC}{dye-sensitized solar cell}
\newacronym{dspvp}{DSPVP}{dye-sensitized photovoltaic photograph}
\newacronym{atr}{ATR}{attenuated total reflectance}
\newacronym{mct}{MCT}{mercury cadmium telluride}
\newacronym{mlct}{MLCT}{metal-to-ligand charge transfer}
\newacronym{ilct}{ILCT}{intraligand charge transfer}
\newacronym{pmma}{PMMA}{polymethylmethyacrylate}
\newacronym{tba}{TBA}{tetrabutylammonium}
\newacronym{gc-ms}{GC-MS}{Gas Chromatography - Mass Spectrometry}
\newacronym{fto}{FTO}{fluorine-doped tin oxide}
\newacronym{pa}{PA}{polyamide}

\chemsetup{greek=upgreek}

\journal{Solar Energy Materials and Solar Cells}

\begin{document}
%%%%%%%%%%%%%%%%%%%%%%%%%%%%%%%%%%%%
% FRONT MATTER
%%%%%%%%%%%%%%%%%%%%%%%%%%%%%%%%%%%%
\begin{frontmatter}

%% Title, authors and addresses
\title{ On the Underlying Mechanism of Light-Induced Patterning of N719-Stained Photoanodes for “Photovoltaic Photographs”} %% Article title

\author[1]{Allyson Robert} %[orcid=0000-0001-7197-9509]

%% Author affiliation
\affiliation[1]{organization={UHasselt, X-LAB},
            addressline={Agoralaan D}, 
            city={Diepenbeek},
            postcode={3590}, 
            country={Belgium}}

% Author affiliation
\affiliation[2]{organization={UHasselt, Analytical \& Circular Chemistry (ACC) Institute for Materials Research (IMO-IMOMEC)},
            addressline={Agoralaan D}, 
            city={Diepenbeek},
%          citysep={}, % Uncomment if no comma needed between city and postcode
            postcode={3590}, 
            country={Belgium}}

% Address/affiliation
\affiliation[3]{organization={UHasselt, Molecular and Physical Plant Physiology},
            addressline={Agoralaan D}, 
            city={Diepenbeek},
%          citysep={}, % Uncomment if no comma needed between city and postcode
            postcode={3590}, 
            country={Belgium}}

\author[1]{Nico Fransaert} %[orcid=0000-0002-4207-7204]
\author[1]{Willem Awouters}
\author[2]{Wouter Marchal}
\author[2]{Peter Adriaensens}
\author[3]{Roland Valcke}
\author[1]{Jean V. Manca} %[orcid=0000-0002-3290-0308]

%%%%%%%%%%%%%%%%%%%%%%%%%%%%%%%%%%%%
% ABSTRACT
%%%%%%%%%%%%%%%%%%%%%%%%%%%%%%%%%%%%
\begin{abstract}
Recently, ``\acrlongpl{pv-photo}'' were proposed as a creative application of \acrlong{pv} technologies, relevant in fields such as architecture or the automotive industry.
In this application an image is created by light-induced patterning of the photoactive layer, causing a local change in the appearance of the solar cell.
In order to further develop this concept, it is crucial to elucidate the mechanism underlying these local changes.
Here, UV-VIS and infrared spectroscopic techniques, as well as \acrlong{gc-ms} and \acrlong{tof-sims}, have been used to investigate the physico-chemical changes induced by this process in the photoactive layer of proof-of-concept photo-patterned \acrlongpl{dssc}.
We show that, for N719 \acrlongpl{pv-photo}, \ch{TiO2} plays a crucial role and that the dye undergoes a multi-step chemical degradation, related to its isothiocyanate ligand.
These insights are of importance for a better understanding of the photo-induced degradation of N719, a more substantiated control of the light-induced patterning process, and to design appropriate light-induced patterning techniques for other classes of solar cells.

\end{abstract}

\begin{graphicalabstract}
\centering
\includegraphics[width=375px,height=150px]{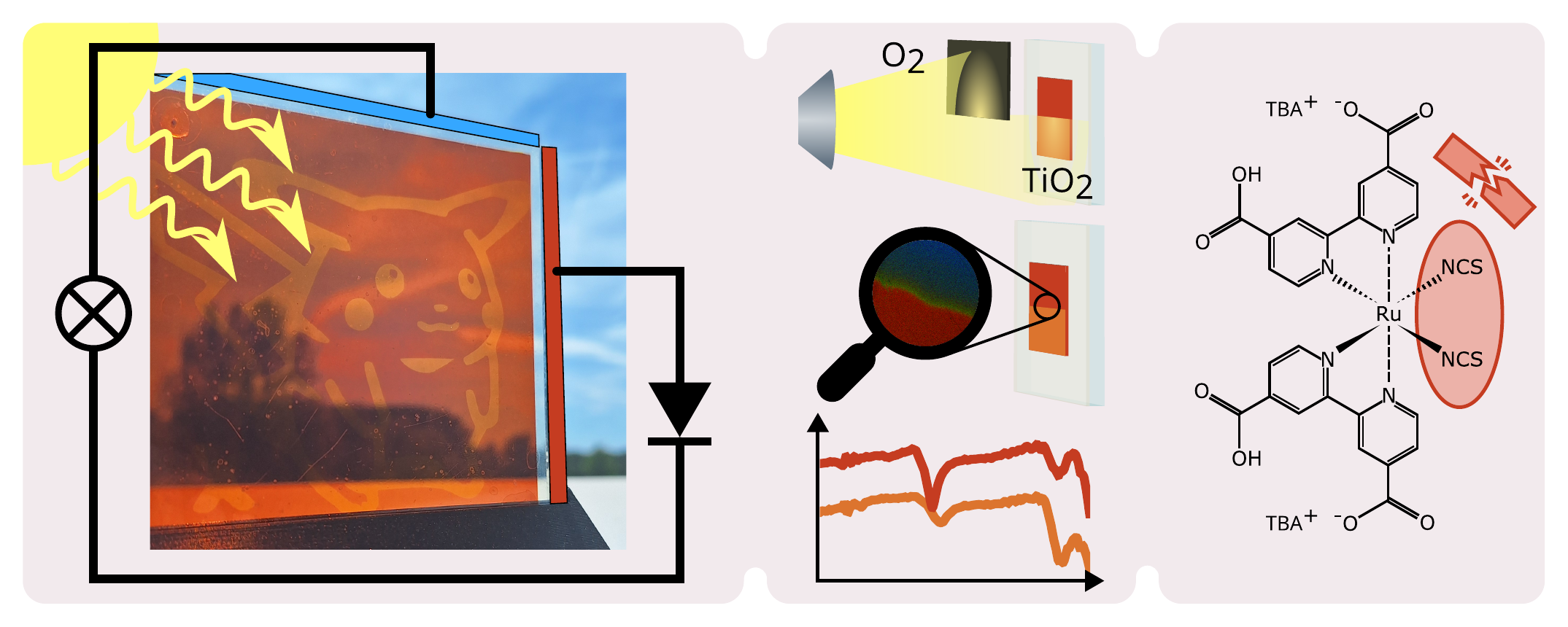}
\end{graphicalabstract}

\begin{highlights}
\item Light-induced patterning of N719 \acrlongpl{dssc} is caused mainly by isothiocyanate degradation
\item Chemical degradation of isothiocyanate in N719 is time dependent and involves multi-step reactions
\item Catalytic effects of \ch{TiO2} are key in the light-induced patterning for photovoltaic photographs based on \acrshort{dssc}
\item Analogous mechanism should be explored to transfer the patterning technique to other classes of \acrshort{pv}
\end{highlights}

\begin{keyword}
photovoltaic photographs \sep N719 \sep photo-induced degradation \sep building-integrated photovoltaics
\end{keyword}

\end{frontmatter}

%%%%%%%%%%%%%%%%%%%%%%%%%%%%%%%%%%%%
% INTRODUCTION
%%%%%%%%%%%%%%%%%%%%%%%%%%%%%%%%%%%%
\section{Introduction}
\label{sec:introduction}
The rapid expansion of \gls{pv} energy production in recent decades has been motivated in part by the challenges posed by climate change, the growing population, and global economic growth \cite{masson_trends_2024,joshi_global_2024,kapsalis_critical_2024,balraj_variational_2022}.
During this period, scientists have kept advancing the technology at an impressive pace by offering new and exciting opportunities.
Some of these new opportunities address the holistic integration of solar installations in areas such as architecture \cite{dambrosio_towards_2021}, agriculture \cite{widmer_agrivoltaics_2024} and even the Internet of Things \cite{farooq_empowering_2024}.
Important for such an integrated approach is to consider freedom of design as well as aesthetics.
Researchers are therefore focusing on properties other than efficiency, such as semi-transparency, flexibility, and colour \cite{shi_semitransparent_2020,hu_agrivoltaics_2024,fara_advanced_2024}.
Installations showcasing some of these applications already exist, such as the International School in Copenhagen \cite{norgaard_perspectives_2019} or the EPFL's SwissTech Convention Center \cite{romande_energie__mediacom_epfls_2013}. 

\begin{figure}
    \centering
    \includegraphics[width=0.7\linewidth]{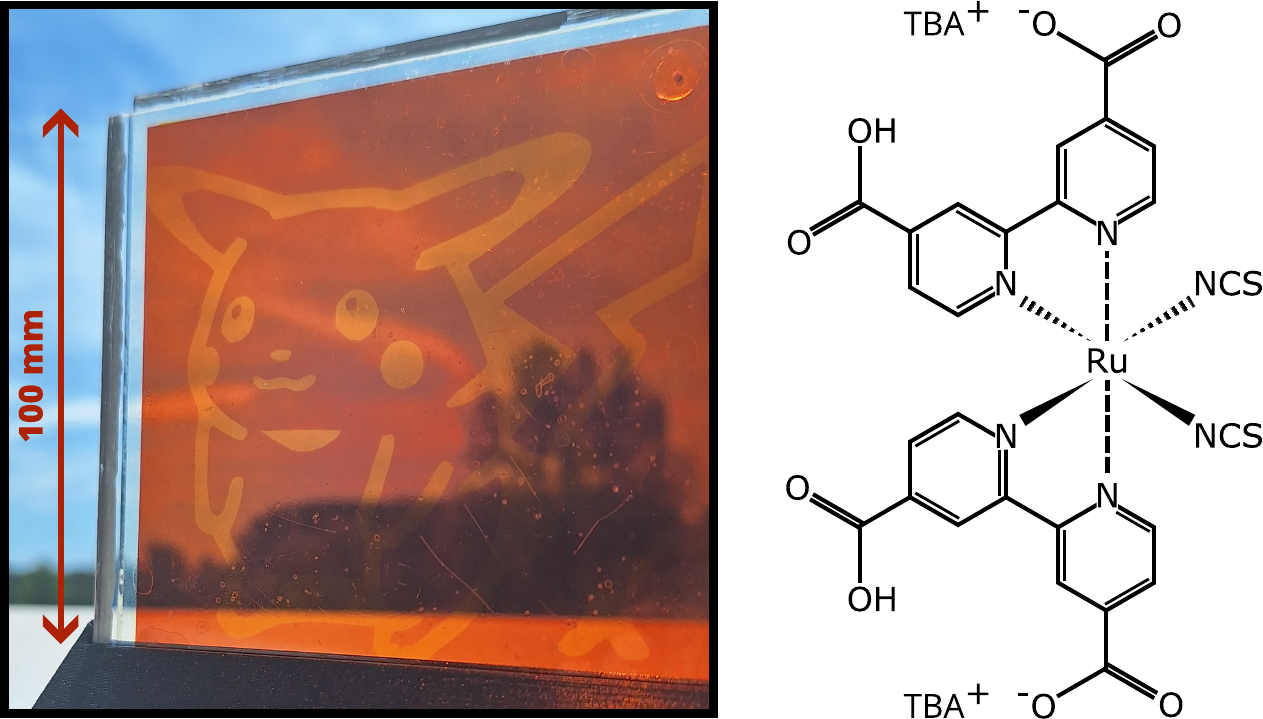}
    \caption{Left: Photograph of a 100 mm x 100 mm ``\acrlong{pv-photo}'', prepared using N719 and containing an image of a popular japanese cartoon figure. Right: Chemical structure of the N719 dye}
    \label{fig:introduction}
\end{figure}

Recently, our group introduced the concept of ``\acrlongpl{pv-photo}'' \cite{hustings_photovoltaic_2022}. 
This innovation allows for spatial control over the visual appearance of solar cells, offering greater flexibility in module design. 
The key to this control lies in patterning the photoactive layer using a light-induced preparation method. 
This technique is straightforward, scalable, cost-effective, and adaptable. 
By integrating design elements directly into energy-producing components, this approach opens new opportunities for applications where both aesthetics and energy generation are essential.
A proof-of-principle of this approach is showcased in \cref{fig:introduction} (left), where an image is embedded into the photo-active layer of a \gls{dssc}.
This class of solar cells is often researched for their advantages in indoor applications \cite{oregan_low-cost_1991,damico_recent_2023}.
Their main working principle is the injection of electrons from a photo-excited dye into the conduction band of titanium oxide (\ch{TiO2}), after which the dye is reduced by a redox-couple.
A typical \gls{dssc} photoanode is made up of a glass substrate coated with a transparent conductive oxide layer, onto which anatase \ch{TiO2} nanoparticles are deposited.
These are then sintered to form a mesoporous structure ensuring conductivity throughout the volume \cite{gratzel_dye-sensitized_2003}.
Embedding a photograph within these cells is achieved by exposing stained photoanodes to light, transmitted selectively through a mask, before applying the counter-electrode.
The key advantages of this technique, over other solar cell patterning approaches such as inkjet printing or mosaic patterning, are design freedom, scalability, and ease of integration into existing production lines \cite{hashmi_dye-sensitized_2016,li_current_2024,mittag_mosaic_2018}.
However, it remained unclear how this photo-induced patterning method causes the visual properties to change and how this is related to the molecular structure of the dye.
Thorough understanding of the patterning mechanism in this proof-of-principle device is crucial to further develop the technique and transfer it to other emerging photovoltaic technologies.
In the initial publication on ``\acrlongpl{pv-photo}'', \citet{hustings_photovoltaic_2022} hypothesised oxygenic photobleaching as an underlying reaction mechanism in the preparation of \acrlongpl{pv-photo}. 
However, the role of the \ch{TiO2} substrate and its effect on the sensitizing dye remained to be investigated explicitly.
An additional research question relates to the illumination times required to achieve the pattern. 
The authors mention the impact of illumination time on the visual contrast and that it differs depending on the photo-stability of the chosen dye.

In this work, we aim to contribute to a better understanding of light-induced patterning of solar cells by examining the specific case of \gls{dssc} \acrshortpl{pv-photo} made using the synthetic ruthenium dye \iupac{di-tetrabutylammonium cis-bis(isothiocyanato)bis(2,2'-bipyridyl-4,4'-dicarboxylato)ruthenium(II)}{}, commonly known in the literature as N719 and whose structure is shown in \cref{fig:introduction} (right).

The following sections discuss how UV-VIS and infrared spectroscopy have been utilised in combination with \gls{tof-sims} to study how the substrate, and short and long illumination times affect N719 in \acrlongpl{pv-photo}.
In doing so, we reveal new insights on the photo-stability of N719 in the presence of \ch{TiO2}.
Additionally, the findings inspire the design of light-induced patterning strategies applicable to other classes of solar cells.

%%%%%%%%%%%%%%%%%%%%%%%%%%%%%%%%%%%%
% MATERIALS AND METHODS
%%%%%%%%%%%%%%%%%%%%%%%%%%%%%%%%%%%%
\section{Materials and experimental methods}
\label{sec:methods}
\subsection{Materials}
\label{sec:methods:materials}
All experiments described here used TechniSolv\textregistered ~ethanol euro denatured (96\%, CE Emb 28007A), purchased from VWR International, as solvent. 
Ruthenizer 535-bisTBA synthetic dye powder, known in the literature as N719, and Ti-Nanoxide T/SP paste (product codes: 21613 and 14451, respectively) were purchased from Solaronix (Switzerland). 
Pre-sintered test cell titania anodes (product code 74111) were also purchased from Solaronix (Switzerland).
These transparent anodes consist of a glass substrate (\SI{20}{\milli\meter} x \SI{20}{\milli\meter}) coated with \gls{fto} onto which an anatase Ti-Nanoxide layer (\SI{6}{\milli\meter} x \SI{6}{\milli\meter}) is homogeneously screen-printed. 
Sodium hydroxide (\ch{NaOH}), used to detach N719 dye from \ch{TiO2} nanoparticles, was purchased from Emplura (pellets, B1835162 131). 
Finally, custom anode holders were 3D printed from neutral BASF Ultrafuse \gls{pa} filament using a Prusa MK4 machine.

\subsection{Time resolved UV-VIS spectroscopy}
\label{sec:methods:tr-uv-vis}
To disentangle the effects of \ch{TiO2} and illumination time on the dye degradation kinetics, two distinct master mixtures of N719 were prepared for \acrlong{tr-uv-vis}.
The first mixture, a \textit{pure solution}, was a \SI{0.3}{\milli\mole\per\liter} solution of powdered N719 in ethanol.
The second mixture, a titanium suspension, was prepared by weighing Ti-Nanoxide T/SP paste and N719 powder to obtain concentrations of \SI{60}{\milli\mole\per\liter} and \SI{0.3}{\milli\mole\per\liter} respectively.
Both mixtures were stirred overnight in the dark, ensuring no exposure to light, and kept in identical glass bottles, sealed by a plastic cap.
After preparation, the mixtures were illuminated concurrently using a Xenon short arc lamp (Thermo Oriel instruments, model 66902, USA), powered by a \SI{150}{\watt} power supply (Oriel instruments, model 6890, USA), at a distance of \SI{30}{\centi\meter} from the lamp opening.
Samples were taken by pipetting \SI{500}{\micro\liter} from the master mixture every 15 minutes during the first two hours of illumination, then every half hour for a total illumination time of 10~hours.
Pure solution samples were then diluted 10-fold before measuring, while the suspension samples required a post-processing step in order to isolate the dye from the \ch{TiO2} nanoparticles (shown in figure S1).
Extraction was achieved through centrifugation (8 minutes at 13000 rpm) using an Eppendorf MiniSpin centrifuge equipped with the F-45-12-11 rotor.
Subsequent removal of the supernatant, addition of \SI{0.1}{\mole\per\liter} \ch{NaOH}, and a final centrifugation cycle \cite{nour-mohhamadi_determination_2005,nour-mohammadi_investigation_2007} yielded samples used for \gls{tr-uv-vis} measurements.
Consequently, the \textit{extracted} solution samples consisted of N719 dissolved in \ch{NaOH} rather than ethanol.
The resulting solution was dark red, leaving behind a white sediment, indicating a successful extraction procedure.
A 5-fold dilution of the extracted solution was used for the UV-VIS measurements. 
These were performed using a DW2000 spectrophotometer from SLM Instruments Inc. (USA) and Olis GlobalWorks software.
The DW2000 spectrometer utilises two distinct lamps for the UV and visible regions of the spectrum and must switch between these.
This transition is seen as a discontinuity in all spectra.
All measurements were done using identical \SI{3}{\milli\liter} \gls{pmma} cuvettes and performed against a pure ethanol reference (for comparison with pure solutions) or \ch{NaOH} reference (for comparison with extracted solutions).

\subsection{Preparation of the ``photovoltaic photographs``}
\label{sec:methods:pv-photo}
\begin{figure}[!htpb]
    \centering
    \includegraphics[width=\linewidth]{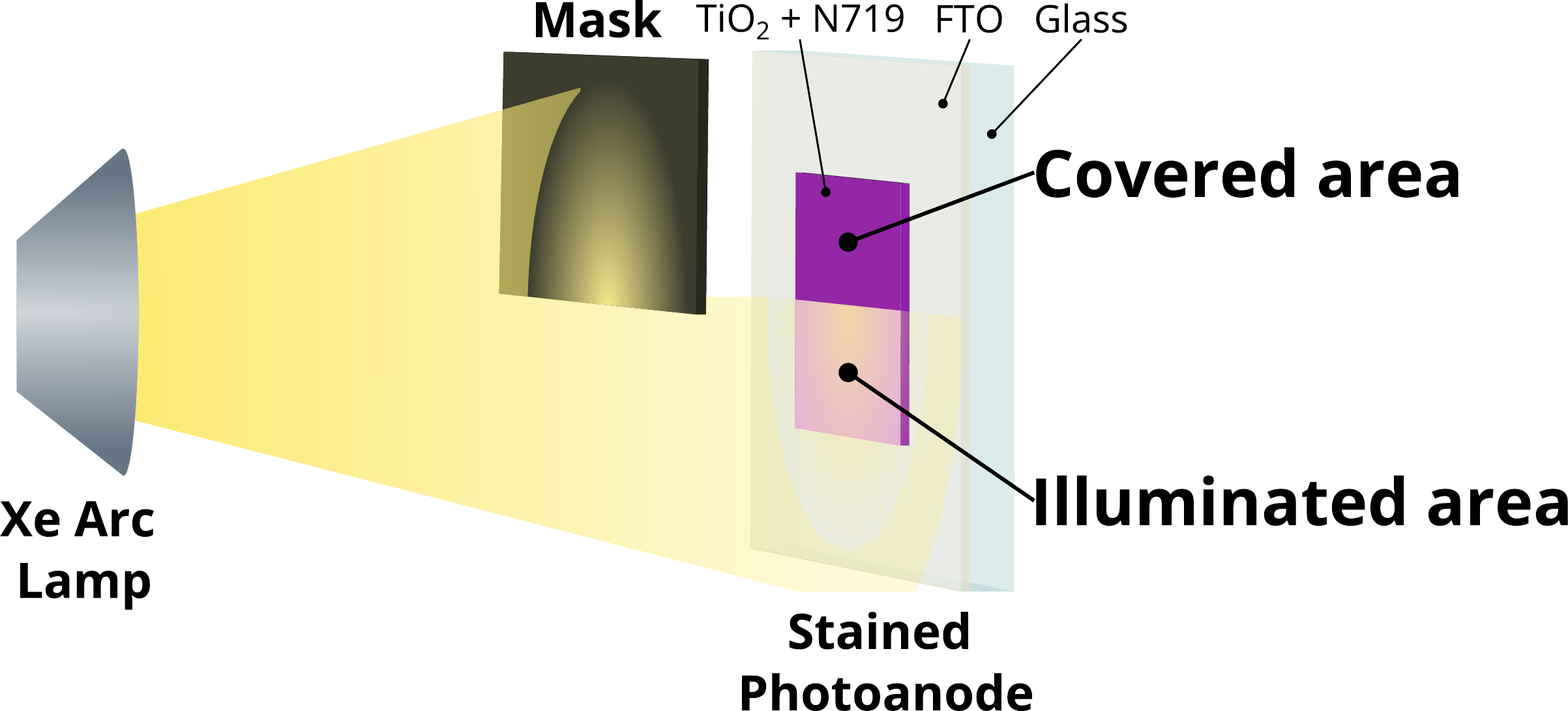}
    \caption{Schematic representation showing the illumination step in the preparation of test \gls{pv-photo} anodes for analysis.
    Light from a Xe short arc lamp was shone onto stained \gls{dssc} photoanodes (consisting of anatase \ch{TiO2} nanoparticles, screenprinted onto \acrshort{fto} covered glass) that were partially covered by a mask.
    The masked area (covered) was minimally exposed to light while the unmasked area (illuminated) showed strong discolouration.}
    \label{fig:illumination-step}
\end{figure}

\Gls{pv-photo} anodes were prepared according to the procedure described by \citet{hustings_photovoltaic_2022}.
Briefly, the purchased \ch{TiO2} anodes were re-fired at \SI{450}{\degreeCelsius} for 30 minutes to remove possible pollutants before being allowed to cool slowly to \SI{60}{\degreeCelsius}.
Subsequently, these anodes were placed in a 3D-printed \gls{pa} holder, with the \ch{TiO2} side facing-down to avoid potential precipitation onto the substrate, fully submerged in an N719 solution prepared freshly as before, and left to stain overnight in the dark.
The next morning, these anodes were removed from the dye solution and rinsed with ethanol to remove residual dye not adsorbed onto the anode and left to air-dry.
After drying, light-induced patterning was achieved by placing the anodes on a glass support in front of a Xenon short arc lamp (Thermo Oriel Instruments, model 66902, USA), which was powered by a \SI{150}{\watt} power supply (Oriel Instruments, model 6890, USA).
The anodes were then illuminated as shown in the diagram of \cref{fig:illumination-step}.
This was carried out under ambient atmosphere, with the \ch{TiO2} layer facing the lamp, ensuring a direct path between the light source and the surface of the sample.

Short (4h) and long (22h) illumination times were chosen to investigate their effect on completed \gls{pv-photo} anodes.
The longer illumination time of 22 hours was chosen for consistency with previous work whereas the 4 hour illumination time ensured a perceptible change while remaining significantly shorter than the reported 22 hour illumination.

In order to simplify the experiments and data analysis, a simple mask was used instead of more intricate, patterned masks.
These ``simplified masks'' consisted of a metal plate placed over a portion of each anode.
Consequently, the prepared \acrlongpl{pv-photo} (figure S4) contained a binary pattern composed of two distinct regions: areas of the anode surface that were masked during illumination, and areas that were directly exposed to light.
These areas of interest will be referred to as \textit{covered} and \textit{illuminated} respectively and are depicted in \cref{fig:illumination-step}.

\subsection{Fourier-transform infrared spectroscopy and gas chromatography - mass spectrometry}
\label{sec:methods:ftir}
To elucidate the chemical changes underlying the light-induced degradation mechanism in N719, semi-transparent \acrshort{dssc} \acrlongpl{pv-photo}, produced as detailed in \cref{sec:methods:pv-photo}, were analysed with \gls{ftir}.
The measurements were performed using a Bruker INVENIO-S spectrometer equipped with a \gls{mct} detector and an \gls{atr} setup equipped with a Germanium crystal.

UV-\acrlong{gc-ms} measurements were carried out with a Frontier UV-1047Xe unit coupled to a EGA/Py-3030D pyrolyzer, a Trace 1300 column GC (Thermo Scientific) and a ISQ7000 MS. 
The pure N719 powder was illuminated with a Xe arc lamp (\SI{280}{\nano\meter}-\SI{450}{\nano\meter}, \SI{300}{\watt}) during 60 minutes at \SI{35}{\degreeCelsius} under dry air atmosphere, whereupon the collected decomposition gasses were analyzed in-situ. 
The sample was subsequently pyrolysed at \SI{550}{\degreeCelsius} and reanalysed to examine the products remaining after the UV exposure.

\subsection{Time-of-flight secondary ion mass spectrometry}
\label{sec:methods:tof-sims}
The effect of the light-induced patterning process on the chemical composition of the photoanode was further analyzed using \acrfull{tof-sims} with a TOF.SIMS 5 instrument (IONTOF GmbH, Münster, Germany) at Imec (Leuven, Belgium). 
Two semi-transparent N719-dyed \ch{TiO_2} anodes were prepared under identical conditions as detailed in \cref{sec:methods:pv-photo} with a light exposure time of either 4 hours or 22 hours.
Since \gls{tof-sims} results from the covered areas were nearly identical, they were combined into a single group. 
Therefore, three groups of areas were analyzed: covered, illuminated for 4 hours, and illuminated for 22 hours. 
Measurements using the so-called high-current bunched mode, which maximizes the mass resolution, were performed with a \SI{30}{\kilo\electronvolt} Bi$_3^+$ primary ion beam transmitting a current of approximately \SI{0.2}{\pico\ampere} at a cycle time of \SI{200}{\micro\second}. 
Areas of {\SI{100}{\micro\meter} $\times$ \SI{100}{\micro\meter}} were analyzed by randomly scanning over $64\times64$\ pixels until reaching a primary ion dose density of $3\times 10^{12}$\,\SI{}{ions/ \centi\meter\squared}. 
A more extensive description of the technical procedure and the data processing strategy have recently been detailed by \citet{fransaert_identifying_2024}.
The fast-imaging mode, optimized for maximum spatial resolution, was additionally employed for the current work. 
A {\SI{250}{\micro\meter} $\times$ \SI{250}{\micro\meter}} area was scanned at $512\times512$\,pixels using a \SI{30}{\kilo\electronvolt} Bi$_3^+$ primary ion beam, transmitting \SI{0.13}{\pico\ampere} at a cycle time of \SI{65}{\micro\second}. 
The total dose density for imaging was $5\times 10^{11}$\,\SI{}{ions/ \centi\meter\squared}.

%%%%%%%%%%%%%%%%%%%%%%%%%%%%%%%%%%%%
% RESULTS
%%%%%%%%%%%%%%%%%%%%%%%%%%%%%%%%%%%%
\section{Results and discussion}
\label{sec:results}

\FloatBarrier
\subsection{\ch{TiO2} plays a crucial role in the light-induced patterning process}
\label{sec:results:tio2}
\begin{figure}[!htpb]
    \centering
    \includegraphics[width=\linewidth]{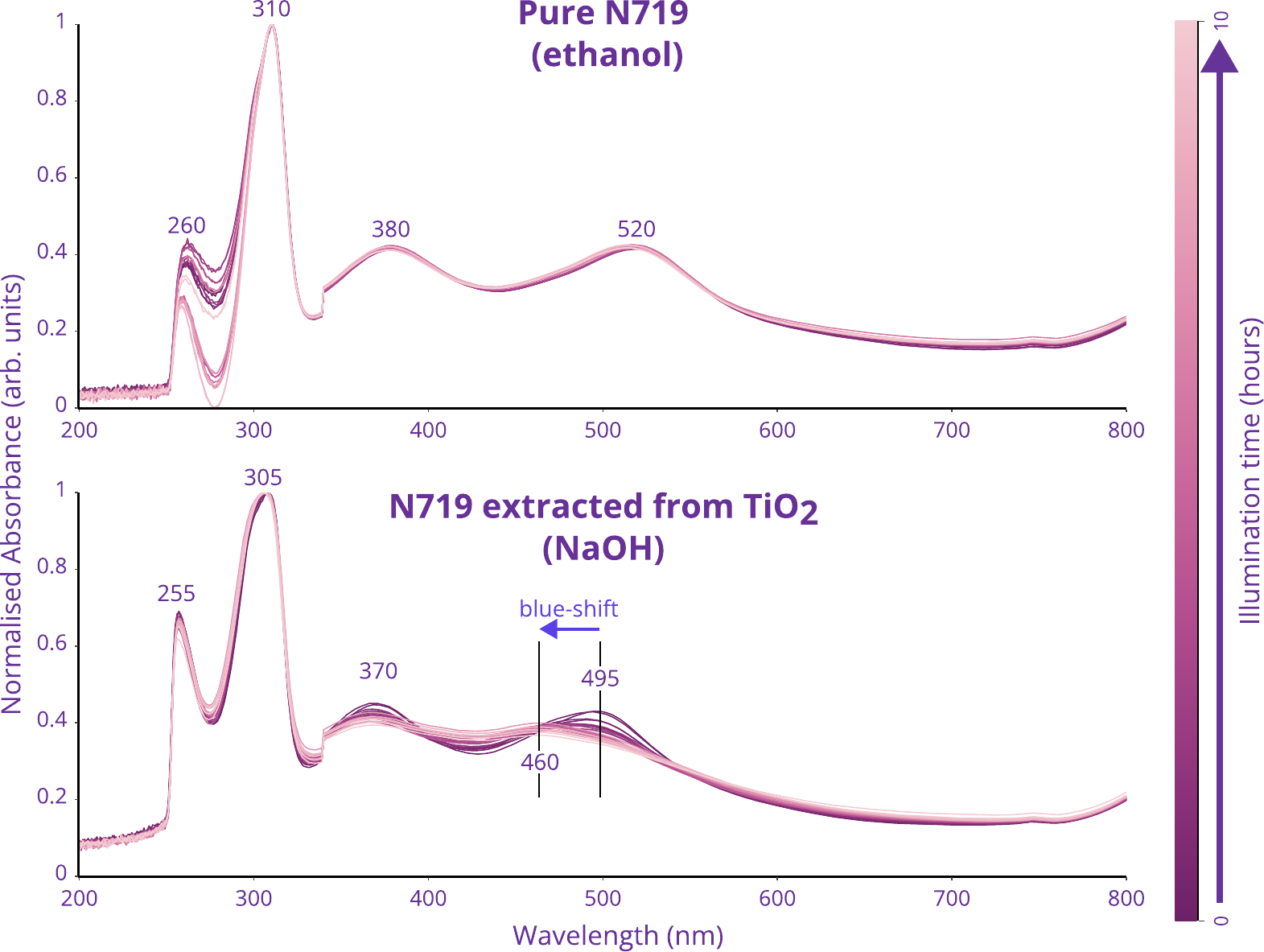}
    \caption{Normalised time resolved UV-VIS spectra for N719 in ethanol (top) and N719 extracted from a suspension of stained \ch{TiO2} nanoparticles (bottom).
    The colour gradient represents illumination time, with a lighter colour corresponding to longer times. 
    The highlighted blue-shift is perceptible as a change of colour of the solution.}
    \label{fig:tr-uv-vis}
\end{figure}

To elucidate the role played by the \ch{TiO2} substrate in the observed visual changes \gls{tr-uv-vis} spectroscopy was employed to compare the spectral changes between pure and surface-bound N719 during illumination.
To this end, two master mixtures were prepared and illuminated concurrently during 10 hours, as described in \cref{sec:methods:tr-uv-vis}.
\Cref{fig:tr-uv-vis} shows the absorbance spectra over time for both the pure (top) and extracted (bottom) solutions, normalised to the height of their \SI{305}{\nano\meter}-\SI{310}{\nano\meter} peak to eliminate uniform changes not dependent on wavelength.

Four prominent peaks are visible in all results.
In the pure solution results these are found around \SI{260}{\nano\meter}, \SI{310}{\nano\meter}, \SI{380}{\nano\meter}, and \SI{520}{\nano\meter}.
The first two peaks in the UV region are associated with \gls{ilct} involving the bipyridine ligands, while the two peaks in the visible region correspond to a \gls{mlct} associated with the isothiocyanate ligands \cite{nazeeruddin_conversion_1993,wen_influence_2012}.

Comparing pure and extracted solutions before illumination reveals differences in these peak locations.
Specifically, the \gls{mlct} peaks are shifted by about \SI{10}{\nano\meter} to \SI{25}{\nano\meter} in the extracted solution results compared to the pure solution spectra.
This initial difference is attributed to the different environments (ethanol vs \ch{NaOH}) used for the UV-VIS measurements (see figure S2).
Over time, both peaks exhibit a drop in intensity as well as significant broadening in the extracted results.
Over the 10~hour measurement period, the peak in the blue-green region, around \SI{495}{\nano\meter}, experiences an additional strong blue-shift to \SI{460}{\nano\meter} on top of this general broadening.
Over the same measurement period this peak only minimally shifted to \SI{515}{\nano\meter}. 
These changes in the visible range are thus greatly accelerated by the presence of \ch{TiO2}.
This observed shift indicates that illumination causes chemical changes to the isothiocyanate ligands coordinated to the ruthenium metal atom.
Specifically, the blue-shift potentially indicates a degradation of these ligands to cyanide \cite{nazeeruddin_conversion_1993}.

\subsection{Chemical degradation of the dye on \ch{TiO2} anodes revealed}
\label{sec:results:ftir}
\begin{figure}[!htbp]
    \centering
    \includegraphics[width=\linewidth]{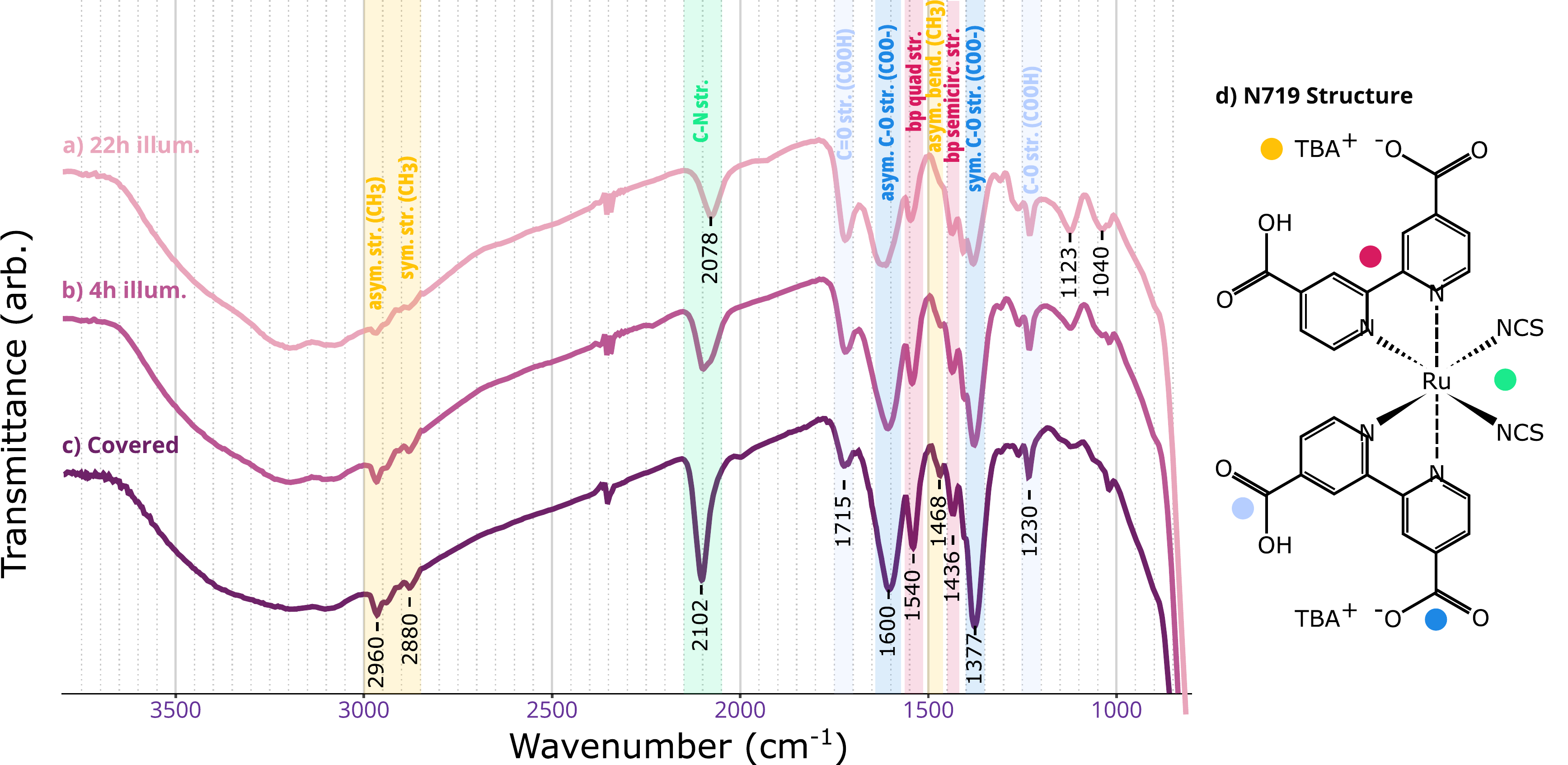}
    \caption{Left: \gls{ftir} spectrum of the illuminated (a and b) and covered (c) areas of anodes prepared according to \cref{sec:methods:ftir}. 
    The spectrum for the covered area is identical to that of a dark area sample (figure S3).
    The coloured regions highlight vibrational bands that were identified using the covered area. 
    Right: Chemical structure of the N719 dye, colour coded to link ligands and functional groups to features in the \gls{ftir} spectrum.}
    \label{fig:ftir}
\end{figure}

Further inquiry into the chemical changes induced in the dye was done by performing \acrlong{ftir} measurements on covered and illuminated anodes, prepared as described in \cref{sec:methods:ftir}.
\Cref{fig:ftir} shows the resulting IR spectra alongside the chemical structure of the N719 dye.

Characteristic vibrational bands were assigned based on the \gls{ftir} spectrum of the negative control anode that was stored in the dark, whose spectra are identical to the covered areas.
These vibrational band were assigned in accordance with the literature \cite{perez_leon_characterization_2006,greijer_agrell_degradation_2003}.

% Discussion of the NCS peak
The \ch{NCS} ligand is probed by the \ch{CN} stretching vibration, initially present at \SI{2102}{\per\centi\meter} but decreasing over time.
Meanwhile, a nearby peak appears under illumination at \SI{2078}{\per\centi\meter}. 
These changes indicate time dependent changes to the \ch{NCS} ligands in the illuminated area, consistent with conclusions found in the literature regarding N719 degradation \cite{nour-mohhamadi_determination_2005,greijer_agrell_degradation_2003}. 

% Protonation of the dye deduced from changes in CO bands
The \ch{C=O} (\SI{1715}{\per\centi\meter}) and \ch{C-O} (\SI{1234}{\per\centi\meter}) vibrational bands of carboxylic acid, as well as the symmetric and asymmetric \ch{C-O} (\SI{1376}{\per\centi\meter}, \SI{1600}{\per\centi\meter}) bands of carboxylate, all showed proportional changes relative to each other. 
In the illuminated area, the carboxylate vibrational bands decreased in intensity relative to the \ch{C=O} band of carboxylic acid. 
Negligible changes in wave number (less than \SI{5}{\per\centi\meter}) are detected for these peaks. 
These observations can be associated with protonation of the dye in the illuminated samples.
% TBA probed by CH2, CH3 vibrations
The bands in the high wavenumber region (\SI{2960}{\per\centi\meter}, \SI{2880}{\per\centi\meter}) as well as the \SI{1470}{\per\centi\meter} peak are typically associated with \ch{CH2} and \ch{CH3} vibrations and have been associated with the \gls{tba} counter-ion \cite{adhikari_anchoring_2024}.
These peaks steadily decrease with illumination time.
Together with the protonation of the dye it can be concluded that this counterion is affected and replaced by hydrogen in the molecule.

% Observations regarding the ring deformations
The vibrational bands assigned to the quadrant stretching and semi-ring deformation of the bipyridine rings (\SI{1540}{\per\centi\meter}, \SI{1436}{\per\centi\meter}), were almost identical in the covered and illuminated areas, suggesting that the bipyridine rings remained unchanged.

% GC-MS shows the products from TBA being degraded
These findings are further supported by \gls{gc-ms} measurements performed on powdered N719 (see tables S1 and S2).
Upon UV illumination, \iupac{tributylamine} was detected as the most prominent volatile degradation product. 
Additional compounds identified in this stage include \iupac{N,N-dibutylformamide}, \iupac{dibutyl disulfide}, \iupac{N,N-dibutylacetamide}, as well as \iupac{butyl (iso)thiocyanate}.
These results confirm the vulnerability of \ch{NCS} while suggesting that the \gls{tba} counter-ion is not only replaced during illumination but also dissociates and readily reacts with the dye’s \ch{NCS} ligands, leading to the formation of various TBA-derived products.

% GC-MS shows bpy stable
No fragments related to the \iupac{bipyridine} ligands were detected in this first measurement. 
However, a second \gls{gc-ms} analysis was conducted after pyrolysis of the remaining sample, in which \iupac{bipyridine}-derived fragments were observed, along with additional \iupac{tributylamine}. 
This confirms the resistance of the \iupac{bipyridine} rings to direct photodegradation and indicates that the \ch{NCS} ligands were eliminated during the initial UV treatment.

% Possible SOx peak
While the frail nature of \ch{NCS} as ligand in the N719 molecule has been reported previously \cite{tsaturyan_ironii_2017}, the fate of the sulfur atom during degradation remains unknown. 
The very strong peak centered at \SI{1126}{\per\centi\meter}, with a shoulder at \SI{1069}{\per\centi\meter}, appearing in the \gls{ftir} data of the illuminated area could provide an additional reaction path, as it has been associated with sulfur oxides in the past \cite{larkin_infrared_2018}.
However, peaks related to isocyanate or nitrile groups would be expected in the \SI{2150}{\per\centi\meter}-\SI{2400}{\per\centi\meter} range and are not present in these results.
% Transition
While \gls{ftir} analysis confirms that patterning primarily involves the changes in the \ch{NCS} ligand, insights into the specific degradation pathway and reaction products involved required additional chemical investigation, which were performed with \acrlong{tof-sims}.

\subsection{TOF-SIMS analysis shows a multi-step degradation pathway for NCS}
\label{sec:results:tof-sims}
\Gls{tof-sims} is a surface-sensitive chemical analysis technique and was used to further examine the process leading to patterning in \acrlongpl{pv-photo}. 
By bombarding the sample's surface with a pulsed primary ion beam, secondary ions are ejected and subsequently detected based on their time of flight, allowing for the construction of mass spectra. 
Here, \gls{tof-sims} was used to analyze the degradation mechanism by comparing three areas of N719-dyed \ch{TiO2} anodes: a covered area, an area illuminated for 4 hours, and another illuminated for 22 hours.

\begin{figure}[!htbp]
    \centering
    \includegraphics[width=0.8\linewidth]{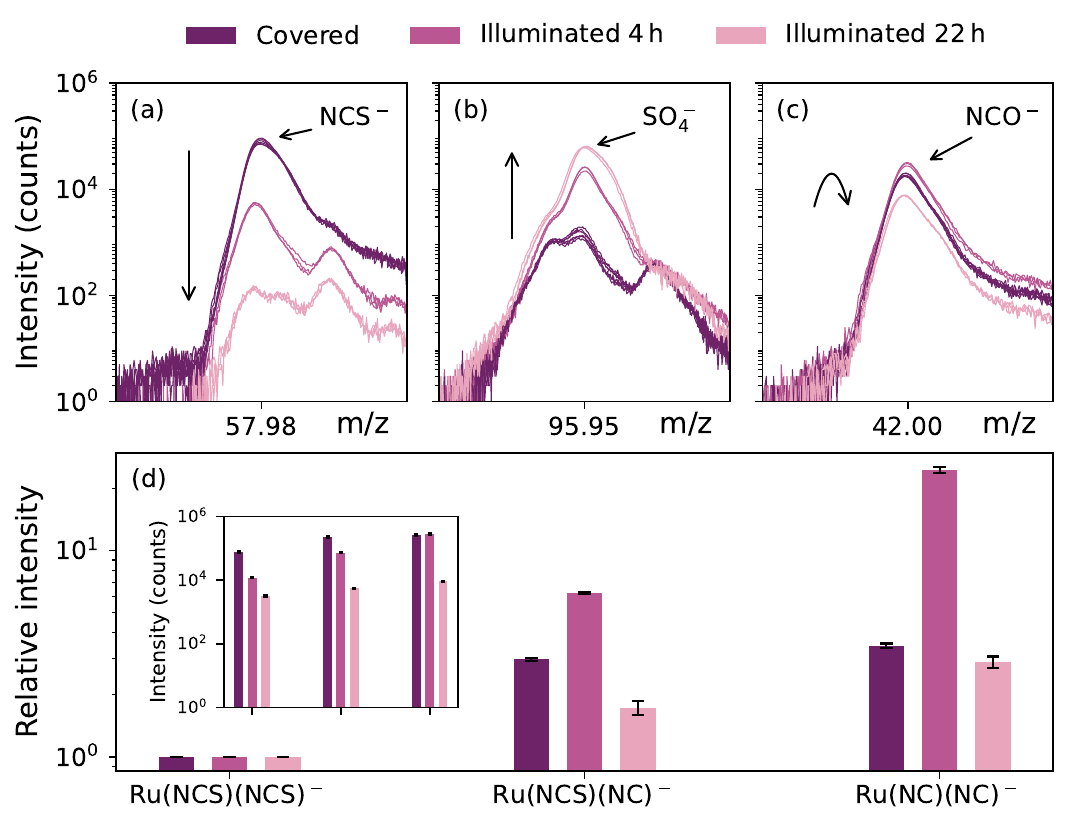}
    \caption{ToF-SIMS comparison of three areas from N719-dyed \ch{TiO_2} anodes: a covered area, an area illuminated for 4 hours, and another illuminated for 22 hours. Spectral intensities are shown for (a) \ch{NCS-}, (b) \ch{SO_4-}, and (c) \ch{NCO-}. The time-dependent nature of the photodegradation is highlighted in (d). The inset shows the absolute peak intensities of \ch{Ru(NCS)(NCS)-}, \ch{Ru(NCS)(NC)-}, and \ch{Ru(NC)(NC)-}, while the main plot presents their relative intensities, calculated as the absolute intensity of each species divided by the absolute intensity of \ch{Ru(NCS)(NCS)-} in the corresponding area. Error bars indicate 95 \% confidence intervals of the means.}
    \label{fig:ToFSIMS-graphs-and-bars}
\end{figure}

\Cref{fig:ToFSIMS-graphs-and-bars} (a) shows a steady decrease in \ch{NCS} ligand intensity during illumination, with near-complete disappearance after 22 hours. 
This irrefutably confirms the substantial photodegradation of the \ch{NCS} ligand, as supported by \gls{ftir} and discussed in previous sections. 
Additionally, \gls{tof-sims} detected sulfur oxide signals, including \ch{SO_4-}, as shown in \cref{fig:ToFSIMS-graphs-and-bars} (b), along with other derivatives like \ch{SO_3-} and \ch{SO_4H-} (not shown here). 
These findings indicate that the primary degradation pathway involves photooxidation of the isothiocyanate sulfur atom, emphasizing oxygen’s role in this process. 

Furthermore, \gls{tof-sims} results suggest that substitution of the \ch{NCS} ligand’s sulfur atom by oxygen may occur. 
\Cref{fig:ToFSIMS-graphs-and-bars} (c) shows an initial increase in \ch{NCO-} concentration after 4 hours of illumination, followed by a decrease after 22 hours. 
This transient increase suggests that substituted compounds also degrade over time, while highlighting the photosensitivity of isothiocyanate's sulfur atom. 
This time-dependent behavior offers crucial insights into the N719 photodegradation mechanism, characterized using a recently proposed \gls{tof-sims} data analysis strategy \cite{fransaert_identifying_2024}. 
This approach identifies compounds that behave differently over the two illumination timescales ($4\,\mathrm{h}$ and $22\,\mathrm{h}$), and highlights variations in the \ch{Ru(NCS)(NCS)} metal ligand's sulfur content. 
Closer examination of peak intensities in \cref{fig:ToFSIMS-graphs-and-bars} (d) reveals that, although the absolute intensities of \ch{Ru(NCS)(NCS)-} and its derivatives decrease, the relative intensities of \ch{Ru(NCS)(NC)-} and \ch{Ru(NC)(NC)-} increase after 4 hours and then decrease after 22 hours. 
These results suggest that sulfur loss occurs rapidly, relative to the overall degradation of the isothiocyanate ligand. 
In combination with previous observations, these time-dependent \gls{tof-sims} results suggest an oxygen-catalyzed, multi-step photodegradation pathway, beginning with sulfur substitution or loss within the \ch{NCS} ligand and culminating in the removal of the ligand.

\begin{figure}[!htbp]
    \centering
    \includegraphics[width=\linewidth]{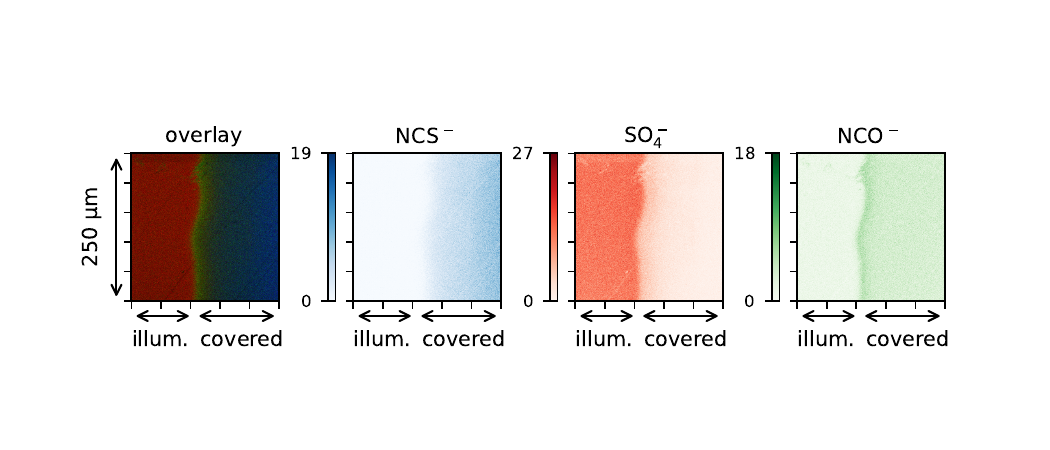}
    \caption{\gls{tof-sims} images of the illumination interface in a N719-dyed \ch{TiO2} anode. The area to the left of the interface was illuminated for 22 hours while the area to the right remained covered. The intensity images for \ch{NCS-} (blue), \ch{SO_4-} (red), and \ch{NCO-} (green) show distinct spatial variations corresponding to the illuminated and covered areas. 
    These images are combined in an RGB overlay to emphasize spatial relationships.}
    \label{fig:ToFSIMS-images}
\end{figure}

Finally, \gls{tof-sims} was used to chemically image the surface composition of a N719-dyed \ch{TiO_2} anode after 22 hours of illumination. 
A $250\times250$\,\SI{}{\micro\meter\squared} region centered around the illumination interface was scanned, producing mass spectra for each pixel.
\Cref{fig:ToFSIMS-images} presents intensity images of \ch{NCS-} (blue), \ch{SO_4-} (red), and \ch{NCO-} (green), both separately and as an RGB overlay. 
As expected, the \ch{NCS-} and \ch{SO_4-} images are anticorrelated, reflecting the degradation of the isothiocyanate ligand and the formation of sulfur oxides. 
The \ch{NCO-} signal is most intense at the illumination interface, likely due to reflective and refractive effects that partially expose this region to light, resulting in substantial degradation over the 22 hour period.
This increase in \ch{NCO-} intensity at the interface mirrors the effect observed in the fully exposed area after 4 hours of illumination, translating the time-dependent degradation process into a spatial variation and reconfirming the multi-step nature of N719 photodegradation during patterning of \acrlongpl{pv-photo}.

%%%%%%%%%%%%%%%%%%%%%%%%%%%%%%%%%%%%
% CONCLUSIONS
%%%%%%%%%%%%%%%%%%%%%%%%%%%%%%%%%%%%
\section{Discussion and conclusion}
\label{sec:conclusion}
While the recently proposed \acrlongpl{pv-photo}, demonstrated using \glspl{dssc}, add to the growing pool of aesthetically oriented applications of solar cells by using colour and semi-transparency to pattern photoanodes, it was not clear which physical and/or chemical changes are responsible for the observed patterning.
In this work, we used various techniques to investigate the occurrence of different mechanisms affecting the dye, triggered by the simultaneous presence of light, oxygen, and \ch{TiO_2}. Specifically, time-resolved UV-VIS spectroscopic measurements reveal a strong and gradual blue-shift in one of the dye’s \acrlong{mlct} peaks.
When the dye is attached to \ch{TiO2} and exposed to light and oxygen, the green peak (\SI{495}{\nano\meter}) in the absorption spectrum quickly shifts by about \SI{35}{\nano\meter} to the blue end of the visual spectrum. 
This indicates a degradation of the \ch{NCS}-ligand in this \ch{Ru}-complex, which was confirmed in all measurements.
These findings are in accordance with previously published results on the degradation of N719 \cite{greijer_agrell_degradation_2003,nour-mohhamadi_determination_2005}.
A drop in intensity in the \gls{ftir} spectra further indicates changes related to the \gls{tba} counter-ion of the dye.
\Gls{gc-ms} measurements confirm that \gls{tba} is affected and additionally show that it reacts with the \ch{NCS} ligands of the dye.
In addition, relative peak intensities of the carboxylate and carboxylic acid groups indicate increasing protonation of the dye with illumination.
These potentially show an impact on the attachment of the dye on the \ch{TiO2} substrate \cite{adhikari_anchoring_2024}.

\Acrlong{tof-sims}, combined with a novel data analysis approach, indicates that the degradation of the \ch{NCS} ligand proceeds through multiple stages rather than as a single-step transformation.
Initially, a steady decrease in \ch{NCS} concentration is observed, along with an increase in \ch{SO4-} levels. 
Concurrently, the intensity of \ch{NCO-} increases at first, then subsequently declines. Supporting evidence from time-resolved UV-VIS and \gls{ftir} measurements suggests that the degradation begins with the rapid loss of sulfur from the \ch{NCS} ligand. 
Some of these sulfur atoms may be replaced by oxygen, forming \ch{NCO-} intermediates, before the ligand is ultimately removed from the dye complex.
Additionally, sulphur atoms can also react with the counter-ion producing dibutyl disulfilde.

These insights contribute to the existing literature on the photo-induced degradation of N719.
Although N719 is extensively studied, to the best of our knowledge, the shown multi-step nature of its photo-induced degradation in the presence of \ch{TiO2} has not been reported before.
%%%
From this work, we have learned that the light-induced patterning of N719-stained photoanodes, used for photovoltaic photographs, arises from mechanisms affecting the dye, triggered by the simultaneous presence of light, oxygen, and \ch{TiO2}. 
These mechanisms lead to observable changes in the absorption spectrum, localized specifically to the illuminated areas. 
These spectral changes can account for the increased perceived transparency and the distinct color shift observed upon visual inspection of the final device. 
Notably, these findings are consistent with the decrease in efficiency reported for complete devices by Hustings et al \cite{hustings_photovoltaic_2022}.

This improved understanding of the light-induced patterning process is of importance for a more controlled design of \acrlongpl{pv-photo} and for a further investigation of their long-term stability.
Further research should use these insights to explore this concept in other classes of solar cells by investigating materials controllably susceptible to (oxygenic) photo-bleaching.

%%%%%%%%%%%%%%%%%%%%%%%%%%%%%%%%%%%%
% BOOKKEEPING
%%%%%%%%%%%%%%%%%%%%%%%%%%%%%%%%%%%%
\section*{CRediT authorship contribution statement}
\textbf{Allyson Robert}: Data curation, Formal analysis, Investigation, Methodology, Visualization, Writing – original draft. 
\textbf{Nico Fransaert}: Data curation, Formal analysis, Investigation, Methodology, Visualization, Writing – original draft.
\textbf{Willem Awouters}: Data curation, Formal analysis, Investigation, Methodology, Writing – review \& editing.
\textbf{Wouter Marchal}: Validation, Writing – review \& editing.
\textbf{Peter Adriaensens}: Validation, Writing – review \& editing.
\textbf{Roland Valcke}: Writing – review \& editing.
\textbf{Jean V. Manca}: Conceptualization, Funding acquisition, Project administration, Supervision, Writing – original draft.

\section*{Declaration of Competing Interest}
The authors declare that they have no known competing financial interests or personal relationships that could have appeared to influence the work reported in this paper.

\section*{Acknowledgments}
The authors thank the colleagues from the X-LAB, T-LAB and OOE research groups at Hasselt University for the numerous inspiring discussions.
Special thanks to M. Van Landeghem for her insightful observations and feedback as well as M. Vanhamel and B. Noppen for performing the \gls{ftir} and UV-\gls{gc-ms} measurements.
This research was supported by Hasselt University's Special Research Fund (BOF22OWB02) and Research Foundation -- Flanders FWO PhD Fellowship grant 11K4324N (N.F.).

\section*{Data availability}
Raw data for all figures shown in this publication are made available through the Zonodo publicly accessible data repository \cite{robert_underlying_2025}.

\section*{Declaration of generative AI and AI-assisted technologies in the writing process}

During the preparation of this work the author(s) used OpenAI's ChatGPT (GPT-4-turbo, April 2025) in order to rephrase specific sentences of the draft for clarity and readability.
After using this tool/service, the author(s) reviewed and edited the content as needed and take(s) full responsibility for the content of the published article.

%%%%%%%%%%%%%%%%%%%%%%%%%%%%%%%%%%%%
% BIBLIOGRAPHY AND APPENDICES
%%%%%%%%%%%%%%%%%%%%%%%%%%%%%%%%%%%%
%% The Appendices part is started with the command \appendix;
%% appendix sections are then done as normal sections

%% For citations use: 
%%       \citet{<label>} ==> Lamport [21]
%%       \citep{<label>} ==> [21]
%%

%% If you have bib database file and want bibtex to generate the
%% bibitems, please use
%%
%%  \bibliographystyle{elsarticle-num-names} 
%%  \bibliography{<your bibdatabase>}

%% else use the following coding to input the bibitems directly in the
%% TeX file.

%% Refer following link for more details about bibliography and citations.
%% https://en.wikibooks.org/wiki/LaTeX/Bibliography_Management

\bibliographystyle{elsarticle-num-names}
\bibliography{paper}

\end{document}